\documentclass[12pt]{article}
\usepackage{graphicx}
\usepackage[cp1251]{inputenc}

\begin{document}
 \noindent {\footnotesize\it
   Astronomy Letters, 2020, Vol. 46, No 6, pp. 370--378.}
 \newcommand{\dif}{\textrm{d}}

 \noindent
 \begin{tabular}{llllllllllllllllllllllllllllllllllllllllllllll}
 & & & & & & & & & & & & & & & & & & & & & & & & & & & & & & & & & & & & & \\\hline\hline
 \end{tabular}

  \vskip 0.5cm
  \centerline{\bf\large Galactic Rotation Parameters Based on Stars from Active}
  \centerline{\bf\large Star-Forming Regions with Data from the Gaia DR2 Catalogue}
   \bigskip
   \bigskip
  \centerline
  {O. I. Krisanova$^{[1]}$\footnote{e-mail: okrisanova@rambler.ru}
   V. V. Bobylev$^{[2]}$ and A. T. Bajkova$^{[2]}$}
   \bigskip

  \centerline{\small\it $^{[1]}$
  St. Petersburg State University, Universitetskaya nab. 7/9, St. Petersburg, 199034 Russia}

  \centerline{\small\it $^{[2]}$ Pulkovo Astronomical Observatory, Russian Academy of Sciences,}

  \centerline{\small\it Pulkovskoe sh. 65, St. Petersburg, 196140 Russia}
 \bigskip
 \bigskip
 \bigskip

 {
{\bf Abstract}---We have studied a sample of more than 25 000 young stars with proper motions and trigonometric
parallaxes from the Gaia DR2 catalogue. The relative errors of their parallaxes do not exceed 10\%.
The selection of stars belonging to active star-forming regions was made by Marton et al. based on
data from the Gaia DR2 catalogue by invoking infrared measurements from the WISE and Planck
catalogues. Low-mass T Tauri stars constitute the majority of sample stars. The following parameters
of the angular velocity of Galactic rotation have been found from them:
 $\Omega_0 =28.40\pm0.11$~km s$^{-1}$ kpc$^{-1}$,
 $\Omega^{'}_0=-3.933\pm0.033$~km s$^{-1}$ kpc$^{-2}$ and
 $\Omega^{''}_0=0.804\pm0.040$~km s$^{-1}$ kpc$^{-3}$. The Oort constants are
$A=15.73\pm0.32$ km s$^{-1}$ kpc$^{-1}$ and $B=-12.67\pm0.34$ km s$^{-1}$ kpc$^{-1}$, while the circular rotation velocity of the solar neighborhood around the Galactic center is $V_0=227\pm4$ km s$^{-1}$ for the adopted Galactocentric distance of the Sun $R_0=8.0\pm0.15$ kpc.
  }


 \subsection*{INTRODUCTION}
Young stars are of great importance for studying the properties of the Galactic disk. Stars of spectral types O and B are youngest among the massive ones; T Tauri stars are youngest among the less massive ones. Such stars are present in active star-forming regions, young open star clusters, and OB and T associations (Blaauw 1964; de Zeeuw et al. 1999; Mel'nik and Dambis 2018).

The Galactic spiral pattern was determined, for example, by Y.M. Georgelin and Y.P. Georgelin (1976)
and Russeil (2003) from the distribution of HII regions and active star-forming regions the photometric
distances to which were estimated from exciting OB stars. Young stars also serve to determine such
parameters of the spiral density wave as the wavelength, the pitch angle, and the velocity perturbation
(Lin et al. 1969; Fern\'andez et al. 2001; Bobylev and Bajkova 2015).

Based on data from various catalogues, the Galactic rotation parameters have been repeatedly determined
both from single and multiple OB stars (Oort 1927; Plaskett and Pearce 1934; Mohr and Mayer 1957; Rubin and Burley 1964; Frogel, and Stothers 1977; Byl and Ovenden 1978; Torra et al. 2000; Bobylev and Bajkova 2015) and from OB associations (Mel'nik et al. 2001; Dambis et al. 2001).

The Gaia DR2 catalogue (Brown et al. 2018; Lindegren et al. 2018) is currently the most important
source of positional and kinematic data. It contains the trigonometric parallaxes and proper motions of
$\sim$1.3 billion stars. For a relatively small fraction of these stars their line-of-sight velocities have been measured. In the Gaia catalogue (Prusti et al. 2016) the photometric measurements are presented in two
broad bands and, therefore, only a very rough classification of stars is possible. For a reliable classification it is necessary to invoke more accurate spectroscopic and photometric data from other sources. Nevertheless, a number of important studies related to the kinematics of various Galactic subsystems have been
performed based on data from the Gaia DR2 catalogue.

Note the compilation by Xu et al. (2018) containing 5772 OB stars with data from the Gaia DR2 catalogue.
The spectral classification of these OB stars was made by various authors from ground-based observations
long before the flight of the Gaia satellite. Bobylev and Bajkova (2019) determined the Galactic
rotation parameters using $\sim$2000 OB stars from the catalogue by Xu et al. (2018) with known parallaxes,
proper motions, and line-of-sight velocities.

A large sample of T Tauri stars containing more than 40 000 stars was produced by Zari et al. (2018).
These stars were selected from the Gaia DR2 catalogue based on kinematic and photometric data, are
located no farther than 500 pc from the Sun, and are closely associated with the Gould Belt. Their spatial
and kinematic properties were studied in detail by Bobylev (2020).

Marton et al. (2019) selected more than 1 million young stellar object candidates. For this purpose,
they combined the Gaia DR2 data with the highly accurate infrared photometry from the WISE experiment
(Wright et al. 2010). Thus, the list of young stars with highly accurate parallaxes and proper motions
from the Gaia DR2 catalogue was expanded significantly. The goal of this paper is to refine the
Galactic rotation parameters based on our sample of
these young stars.

\begin{figure}[t]
{\begin{center}
   \includegraphics[width=0.95\textwidth]{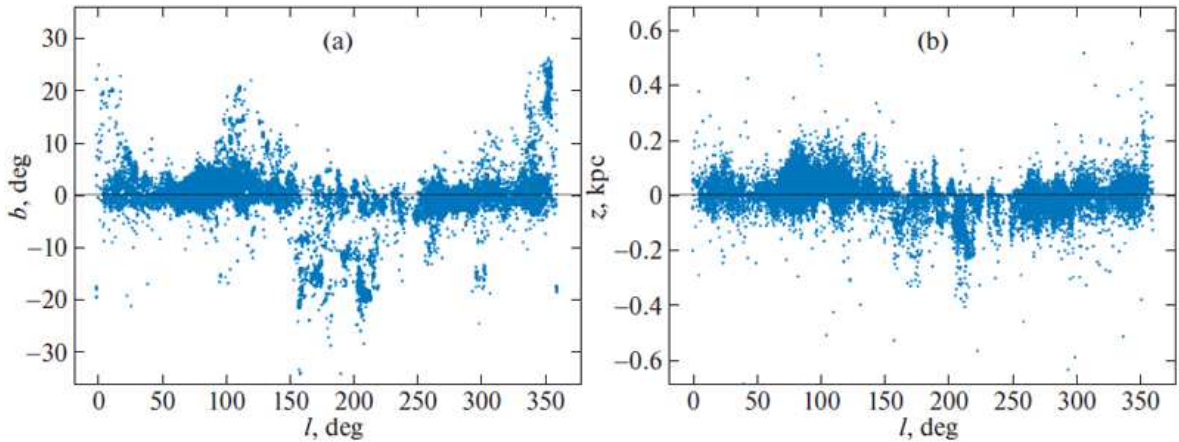}
 \caption{
The distributions of stars on the $l-b$ (a) and $l-z$ (b) planes.
  } \label{f-lb}
\end{center}}
\end{figure}
\begin{figure}[t]
{\begin{center}
   \includegraphics[width=0.5\textwidth]{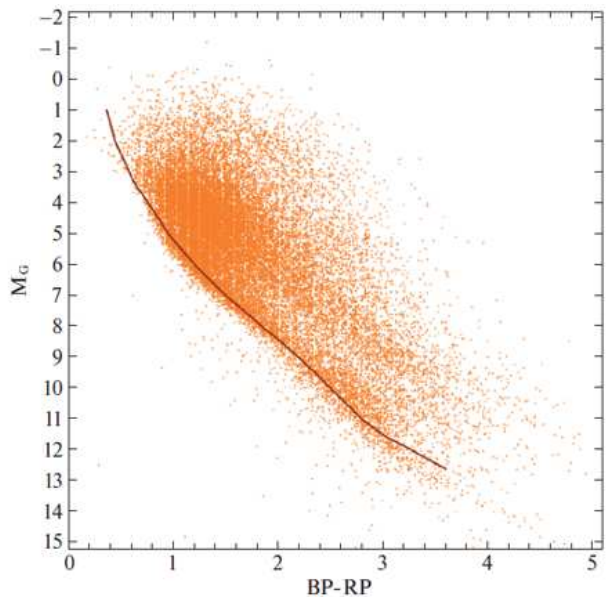}
 \caption{
The color–absolute magnitude diagram constructed from stars with relative parallax errors less than 10\%. The solid line marks the main sequence.
  } \label{f-GR}
\end{center}}
\end{figure}

 \subsection*{DATA}
In this paper we use the selection of young Galactic stellar objects from Marton et al. (2019). The data
were obtained from a combination of orbital observations on board the WISE and Gaia satellites and,
therefore, this database is called GaiaDR2xAllWISE.

The WISE infrared space telescope was launched by NASA into a near-Earth orbit in 2009. Its objective
was an all-sky survey in four infrared bands (3.3, 4.7, 12, and 23 $\mu$m). The AllWISE online
electronic database was created for a convenient use
of the WISE observations (Cutri et al. 2013).

Young stars are located predominantly in regions containing a large amount of dust. A detailed Galactic
extinction map was constructed from infrared observations with the Planck space telescope (Adam
et al. 2016). To find out how a source is associated with a dust region, Marton et al. (2019) used the dust
opacity ($\tau$) for each object from this map.

The Gaia DR2 catalogue contains the trigonometric parallaxes, proper motions, and photometric
data for $\sim$1.3 billion stars. The mean trigonometric parallax errors lie in the range 0.02--0.04 mas for
bright stars (G$<$15$^m$) and reach 0.7 mas for faint stars (G=20$^m$). For more than 7 million stars of
spectral types F--G--K the line-of-sight velocities were determined with a mean error $\sim$1 km s$^{-1}$. The
line-of-sight velocities from ground-based spectroscopic measurements are used in analyzing the space
velocities of younger stars, for example, OB stars.

In total, the catalogue by Marton et al. (2019) produced by combining the Gaia DR2, WISE, and
Planck measurements contains more than 100 million objects of different nature. For a detailed study of the young Milky Way population they compiled a list of more than 1.1 million young stellar object
candidates. As these authors point out, the Gaia satellite detected a number of highly variable sources
In many cases, the physical nature of such sources is unknown, but $\sim$30\% of them are most likely associated with the flare activity of young stars.

The catalogue by Marton et al. (2019) proper contains AllWISE and 2MASS photometric information,
Gaia DR2 ID, and the probability for a star to belong to one of the four classes under consideration.
These authors introduced four main classes: young stellar objects (YSOs), extragalactic objects, main-sequence
stars, and evolved stars.

We took the parallaxes, proper motions, and line-of-sight velocities independently from the Gaia DR2
catalogue. For this purpose, we used the ARI's Gaia TAP Service
 (http://gaia.ari.uni-heidelberg.de/tap.html).

Our working sample was produced by extracting all of the stars with a probability of being a YSO more
than 80\% from the catalogue by Marton et al. (2019). There were 31 937 and 25 508 such stars with relative
trigonometric parallax errors less than 15\% and 10\%, respectively.

In this paper we investigate the stars with relative trigonometric parallax errors less than 10\%. Figure 1
presents the distributions of 25 508 stars on the celestial sphere and the $l-z$ plane. On the one hand,
some of the stars have fairly high Galactic latitudes, as can be seen from Fig. 1a. This is important for
a reliable determination of such parameters as, for example, the velocity $W_\odot$. On the other hand, it can
be seen (Fig. 1b) that all stars are highly concentrated to the Galactic plane; their heights $z$ usually do not exceed 200 pc.

Figure 2 presents the Hertzsprung–Russell diagram constructed from 25 508 stars using photometric
data from the Gaia DR2 catalogue without any correction for interstellar extinction and reddening.
It can be clearly seen from Fig. 2 than the bulk of the sample stars occupy an extensive region above
the main sequence typical for young T Tauri stars. A similar picture can be seen in Fig. 5 from Zari
et al. (2018), where T Tauri stars were selected from the Gaia DR2 catalogue based on their kinematic
characteristics.

 \subsection*{METHODS}
 \subsubsection*{Galactic Rotation Parameters}
From observations we know three stellar velocity components: the line-of-sight velocity $V_r$ and the two tangential velocity components $V_l=4.74r\mu_l\cos b$ and $V_b=4.74r\mu_b$ along the Galactic longitude $l$ and latitude $b,$ respectively, expressed in km s$^{-1}$ are known from observations. Here, the coefficient 4.74 is the ratio of the number of kilometers in an astronomical unit to the number of seconds in a tropical year. The proper motion components $\mu_l\cos b$ and $\mu_b$ are expressed in mas yr$^{-1}$.

To determine the parameters of the Galactic rotation curve, we use the equations derived from Bottlinger's
formulas in which the angular velocity $\Omega$ was expanded in a series to terms of the second order of
smallness in $r/R_0$:
\begin{equation}
 \begin{array}{lll}
 V_r=-U_\odot\cos b\cos l-V_\odot\cos b\sin l-W_\odot\sin b\\
 +R_0(R-R_0)\sin l\cos b\Omega^\prime_0+0.5R_0(R-R_0)^2\sin l\cos b\Omega^{\prime\prime}_0,
 \label{EQ-1}
 \end{array}
 \end{equation}
 \begin{equation}
 \begin{array}{lll}
 V_l= U_\odot\sin l-V_\odot\cos l-r\Omega_0\cos b\\
 +(R-R_0)(R_0\cos l-r\cos b)\Omega^\prime_0+0.5(R-R_0)^2(R_0\cos l-r\cos b)\Omega^{\prime\prime}_0,
 \label{EQ-2}
 \end{array}
 \end{equation}
 \begin{equation}
 \begin{array}{lll}
 V_b=U_\odot\cos l\sin b + V_\odot\sin l \sin b-W_\odot\cos b\\
 -R_0(R-R_0)\sin l\sin b\Omega^\prime_0-0.5R_0(R-R_0)^2\sin l\sin b\Omega^{\prime\prime}_0,
 \label{EQ-3}
 \end{array}
 \end{equation}
where $R$ is the distance from the star to the Galactic rotation axis (cylindrical radius):
  \begin{equation}
 R^2=r^2\cos^2 b-2R_0 r\cos b\cos l+R^2_0.
 \end{equation}
The quantity $\Omega_0$ is the angular velocity of Galactic rotation at the solar distance $R_0,$ the parameters $\Omega^{\prime}_0$ and $\Omega^{\prime\prime}_0$ are the corresponding derivatives of the angular velocity, and $V_0=|R_0\Omega_0|.$ In Eqs. (1)--(3) six unknowns are to be determined: 
 $U_\odot,$ $V_\odot,$ $W_\odot,$ $\Omega_0,$ $\Omega^\prime_0$ and $\Omega^{\prime\prime}_0$.
The Oort constants $A$ and $B$ are also of interest. Their values can be found from the following
expressions:
\begin{equation}
 A=0.5\Omega^{\prime}_0R_0,\quad
 B=-\Omega_0+A. \label{AB}
\end{equation}

The kinematic parameters are determined by solving the conditional equations (1)--(3) by the
least-squares method (LSM). We use weights of the form  $w_r=S_0/\sqrt {S_0^2+\sigma^2_{V_r}},$
 $w_l=S_0/\sqrt {S_0^2+\sigma^2_{V_l}}$ and $w_b=S_0/\sqrt {S_0^2+\sigma^2_{V_b}},$
where $S_0$ is the ``cosmic'' dispersion, $\sigma_{V_r}$ and $\sigma_{V}$ are the dispersions of the
corresponding observed velocities. $S_0$ is comparable to the root-mean-square residual $\sigma_0$ (the error per unit weight) that is calculated when solving the conditional equations (1)--(3). In this paper the
adopted values of $S_0$ lie in the range 15--20 km s$^{-1}$. The system of equations (1)--(3) is solved in several iterations using the 3$\sigma$ criterion to eliminate the stars with large residuals.

 \subsubsection*{Choosing $R_0$}
At present, a number of studies devoted to determining the mean distance from the Sun to the
Galactic center using its individual determinations in the last decade by independent methods have
been performed. For example, $R_0=8.0\pm0.2$ kpc (Vall\'ee 2017), $R_0=8.4\pm0.4$ kpc (de Grijs and
Bono 2017), or $R_0=8.0\pm0.15$ kpc (Camarillo et al. 2018).

Note also some of the first-class individual measurements of this quantity made in recent years. Having
analyzed a 16-year-long series of observations of the motion of the star S2 around the supermassive
black hole at the Galactic center, Abuter et al. (2019) found $R_0=8.178\pm0.022$ kpc. Based on an independent
analysis of the orbit of the star S2, Do et al. (2019) found $R_0=7.946\pm0.032$ kpc. Based on masers from the Japanese VERA Program, Hirota et al. (2020) obtained an estimate of $R_0=7.9\pm0.3$ kpc. Based on the listed results, in this paper we adopt $R_0=8.0\pm0.15$ kpc.

\begin{figure}[t]
{\begin{center}
   \includegraphics[width=0.5\textwidth]{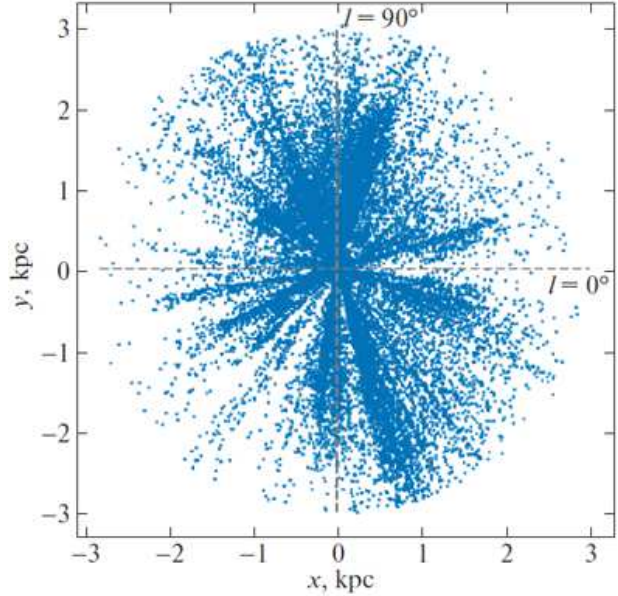}
 \caption{
The distribution of stars with relative trigonometric parallax errors less than 10\% on the Galactic $x-y$
plane. The Sun lies at the coordinate origin; the Galactic center is on the right.
  } \label{f-xy}
\end{center}}
\end{figure}
\begin{figure}[t]
{\begin{center}
   \includegraphics[width=0.7\textwidth]{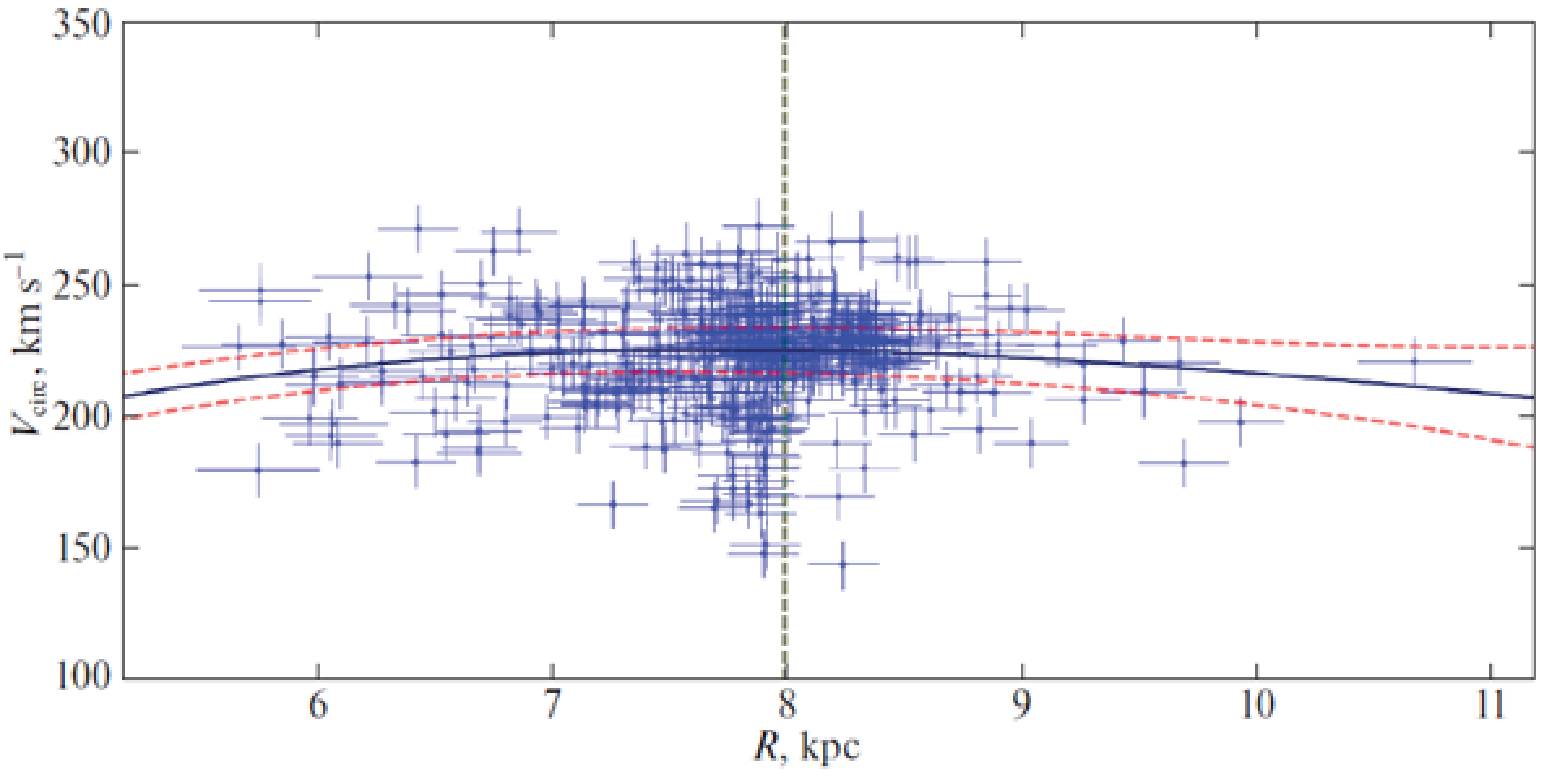}
 \caption{
The Galactic rotation curve constructed from 466 stars withmeasured line-of-sight velocities. The vertical dashed line marks the Sun’s position; the boundaries of the 1$\sigma$ confidence region are indicated.
  } \label{f-rotation}
\end{center}}
\end{figure}

 \subsection*{RESULTS}
Figure 3 presents the distribution of stars from our working sample on the Galactic $x-y$ plane. In
addition to the constraint on the parallax error (10\%), we also use the constraint on the distance $r<3$ kpc.
As follows from this figure, no connection with the spiral structure is visible.

As a result of the LSM solution of the system of two equations (2) and (3), from the proper
motions of 25 508 stars within 3 kpc of the Sun with relative trigonometric parallax errors less than
10\% we found the components of the group velocity vector $(U_\odot,V_\odot,W_\odot)=(10.69,14.49,7.67)\pm(0.14,0.24,0.11)$ km s$^{-1}$ and the following components
of the angular velocity of Galactic rotation:
 \begin{equation}
 \label{sol-1}
 \begin{array}{lll}
       \Omega_0=~28.63\pm0.12~\hbox{km s$^{-1}$ kpc$^{-1}$},\\
   \Omega^{\prime}_0=-4.019\pm0.036~\hbox{km s$^{-1}$ kpc$^{-2}$},\\
  \Omega^{\prime\prime}_0=~0.796\pm0.044~\hbox{km s$^{-1}$ kpc$^{-3}$}.
 \end{array}
 \end{equation}
In this solution the error per unit weight is $\sigma_0=17.0$ km s$^{-1}$, the Oort constants are $A=16.08\pm0.33$ km s$^{-1}$ kpc$^{-1}$ and $B=-12.55\pm0.35$ km s$^{-1}$ kpc$^{-1}$, and the linear rotation velocity of the solar neighborhood around the Galactic center is $V_0=229\pm4$ km s$^{-1}$.

A problem with the Gaia DR2 trigonometric parallaxes is known since the publication of the Gaia DR2 catalogue or, more specifically, a correction $\Delta\pi$ from $-$0.03 to $-$0.05 mas is needed (Lindegren et al. 2018; Arenou et al. 2018). Taking into account the determination of this correction in Yalyalieva et al. (2018), Riess et al. (2018), and Zinn et al. (2018), we must add a correction of 0.050 mas to all the original stellar parallaxes from the Gaia DR2 catalogue, i.e., $\pi_{new}=\pi+0.05$ mas. The solution (6) was obtained fromthe original parallaxes, but we obtain all of the succeeding solutions with the corrected parallaxes.

Thus, as a result of the LSM solution of the system of two equations (2) and (3) with the same
constraints as those used in seeking the solution (6), but applying the correction to the stellar parallaxes,
we found the components of the group velocity vector $(U_\odot,V_\odot,W_\odot)=(9.99,14.04,7.25)\pm(0.13,0.22,0.10)$ km s$^{-1}$ and the following components
of the angular velocity of Galactic rotation:
 \begin{equation}
 \label{sol-2}
 \begin{array}{lll}
      \Omega_0=~28.40\pm0.11~\hbox{km s$^{-1}$ kpc$^{-1}$},\\
  \Omega^{\prime}_0=-3.933\pm0.033~\hbox{km s$^{-1}$ kpc$^{-2}$},\\
 \Omega^{\prime\prime}_0=~0.804\pm0.040~\hbox{km s$^{-1}$ kpc$^{-3}$}.
 \end{array}
 \end{equation}
In this solution the error per unit weight is $\sigma_0=16.0$ km s$^{-1}$, the Oort constants are 
$A=15.73\pm0.32$ km s$^{-1}$ kpc$^{-1}$ and $B=-12.67\pm0.34$ km s$^{-1}$ kpc$^{-1}$, and the linear rotation velocity of the solar neighborhood around the Galactic center is $V_0=227\pm4$ km s$^{-1}$.

 \begin{table}[t]
 \caption[]{\small
The Galactic rotation parameters found from 466 stars with measured line-of-sight velocities and relative
trigonometric parallax errors less than 10\%.
 }
  \begin{center}  \label{t:01}  \small
  \begin{tabular}{|l|r|r|r|r|r|}\hline
   Parameters   & $V_l+V_b+V_r$ &    $V_l+V_b$ &        $V_r$ \\\hline
 $U_\odot,$ km s$^{-1}$ &  $ 9.8\pm0.9$ & $10.8\pm1.2$ & $ 8.8\pm1.8$ \\
 $V_\odot,$ km s$^{-1}$ &  $15.3\pm1.0$ & $13.7\pm1.7$ & $18.0\pm1.5$ \\
 $W_\odot,$ km s$^{-1}$ &  $ 8.1\pm0.8$ & $ 8.7\pm0.9$ & --- \\

  $\Omega_0,$  km s$^{-1}$ kpc$^{-1}$     & $28.1\pm0.9$  & $ 29.5\pm1.2$  & ---  \\
  $\Omega^{'}_0,$ km s$^{-1}$ kpc$^{-2}$  & $-3.74\pm0.20$ & $-3.98\pm0.31$ & $-3.83\pm0.36$ \\
  $\Omega^{''}_0,$ km s$^{-1}$ kpc$^{-3}$ & $0.49\pm0.25$ & $ 0.48\pm0.32$ & $ 0.87\pm0.57$ \\
   $\sigma_0,$    km s$^{-1}$             &         17.2   &          18.6  &           20.5 \\
  \hline
 \end{tabular}\end{center} \end{table}

Table 1 presents the results of our kinematic analysis of a sample of 466 stars for which both proper
motions and line-of-sight velocities are available. We took only those stars whose line-of-sight velocity
measurement errors did not exceed 10 km s$^{-1}$, with relative parallax errors less than 10\%, and with distances to them less than 3 kpc. Table 1 gives the Galactic rotation parameters derived by three methods.
The second column of the table presents the results of the LSM solution in seeking which we solved
all three conditional equations (1)--(3); this solution is designated as ``$V_l+V_b+V_r$''. In the third column we solved two conditional equations, (2) and (3); this solution is designated as ``$V_l+V_b$''. In the last column we solved one conditional equation (3), and this solution is designated as ``$V_r$''. As can be seen from Eq. (1), the angular velocity of Galactic rotation cannot be
determined only from an analysis of the line-of-sight
velocities. In addition, the velocity $W_\odot$ is poorly
determined with this approach and, therefore, here we
took it to be 7 km s$^{-1}$.

A comparison of the values of $\Omega_0'$ found by various methods allows the distance scale factor $p$ to
be determined (Zabolotskikh et al. 2002; Rastorguev et al. 2017; Bobylev and Bajkova 2017a); in our case,
$p=(\Omega^{'}_0)_{V_r}/(\Omega^{'}_0)_{V_l}$. This method is based on the fact that the error in $\Omega^{'}_0)_{V_l}$  depends very weakly on the distances to the stars used.

We can estimate p by two methods: (1) based on the solutions ``$V_l+V_b$'' and ``$V_r$' from Table 1, we
find $p=(-3.83)/(-3.98)=0.96\pm0.12$; (2) from the combination of the solution Vr from Table 1 with the
solution (7), where the error in $\Omega'_{0V_l}$ is small, we obtain $p=(-3.83)/(-3.933)=0.97\pm0.09$. Here,
we calculated the error in p based on the relation 
 $\sigma^2_p=(\sigma_{\Omega'_{0V_r}}/\Omega'_{0V_l})^2+
     (\Omega'_{0V_r}\cdot\sigma_{\Omega'_{0V_l}}/\Omega'^2_{0V_l})^2.$ 
Having analyzed more than 50 000 stars from the TGAS catalogue (Brown et al. 2016), Bobylev and Bajkova
(2018) obtained an estimate of $p=0.97\pm0.04$ by this method. From open star clusters with
data from the Gaia DR2 catalogue Bobylev and Bajkova (2018b) found $p=1.00\pm0.04.$ We can conclude that the distance scale factor is close to unity and, therefore, the distances do not require a correction factor.

Figure 4 plots the circular velocities $V_{circ}$ against distance $R$ for 466 stars with measured line-of-sight velocities. The Galactic rotation curve was constructed
with the parameters given in the second
column of Table 1. The boundaries of the confidence
region were found by the Monte Carlo method.

 \begin{table}[t]
 \caption[]{\small
The group velocity vector and the Oort constants
 }
  \begin{center}  \label{UVW}   \small
  \begin{tabular}{|c|c|c|c|c|c|c|}\hline
     $U_\odot,$ &  $V_\odot,$    &    $W_\odot,$   &    $A,$    &    $B,$    & Ref & Sample \\
    km s$^{-1}$ &  km s$^{-1}$   &  km s$^{-1}$    & km s$^{-1}$ kpc$^{-1}$ &  km s$^{-1}$ kpc$^{-1}$  & (*) & \\\hline
 $ 7.5 \pm1.9 $ & $11.2 \pm1.3 $ &           --- & $17.8\pm1.5$   & $-12.5\pm1.7$   & (1) & OB stars     \\
 $ 6.5 \pm1.8 $ & $12.1 \pm1.7 $ & $7.2 \pm1.2 $ & $17.0\pm1.5$   & $-10.5\pm2.0$   & (2) & Ceph.+OSCs \\
 $ 7.9 \pm0.7 $ & $11.7 \pm0.8 $ & $7.4 \pm0.6 $ & $16.20\pm0.38$ & $-12.64\pm0.51$ & (3) & Cepheids      \\
 $11.1 \pm1.3 $ & $18.3 \pm1.2 $ & $8.8 \pm1.1 $ & $16.63\pm0.38$ & $-13.30\pm0.65$ & (4) & 130 masers    \\  $8.53\pm0.38$ & $11.22\pm0.46$ & $7.83\pm0.32$ & $16.40\pm0.23$ & $-12.31\pm0.32$  & (5) & OSCs         \\
 $ 6.53\pm0.24$ & $ 7.27\pm0.31$ &           --- & $16.14\pm0.13$ & $-13.56\pm0.17$ & (6) & OB stars      \\
 \hline
 \end{tabular}\end{center}
 {\small\protect (1) Melnik et al. (2001); (2) Zabolotskikh et al. (2002); (3) Bobylev (2017); (4) Rastorguev et al. (2017); (5) Bobylev and Bajkova (2019b); (6) Bobylev and Bajkova (2019a).
 }
  \end{table}
  \begin{table}[t]
 \caption[]{\small
Galactic rotation parameters
 }
  \begin{center}  \label{Omega}   \small
  \begin{tabular}{|c|c|c|c|c|c|c|}\hline
 $\Omega_0,$ & $\Omega^{\prime}_0,$ & $\Omega^{\prime\prime}_0,$ & $R_0,$ & $V_0,$ & Ref & Sample \\
 km s$^{-1}$ kpc$^{-1}$ & km s$^{-1}$ kpc$^{-2}$  & km s$^{-1}$ kpc$^{-3}$  &  kpc   & km s$^{-1}$ &    &   \\\hline

 $30.2 \pm0.8 $ & $-5.0 \pm0.2 $ & $1.5  \pm0.2  $ & $7.1\pm0.5 $ & $214\pm16$ & (1) & OB stars    \\
 $27.47\pm1.39$ & $-4.54\pm0.24$ & $1.09 \pm0.19 $ & $7.5\pm0.5 $ & $206\pm14$ & (2) & Ceph.+OSCs \\
 $28.84\pm0.33$ & $-4.05\pm0.10$ & $0.805\pm0.067$ & $8.0\pm0.2 $ & $231\pm6$  & (3) & Cepheids \\
 $28.93\pm0.53$ & $-3.96\pm0.07$ & $0.87 \pm0.03 $ & $8.4\pm0.12$ & $243\pm6$  & (4) & 130 masers    \\
 $28.71\pm0.22$ & $-4.10\pm0.06$ & $0.736\pm0.033$ & $8.0\pm0.15$ & $230\pm5$  & (5) & OSCs  \\
 $29.70\pm0.11$ & $-4.03\pm0.03$ & $0.620\pm0.014$ & $8.0\pm0.15$ & $238\pm5$  & (6) & OB stars \\
 $28.96\pm0.27$ &      ---   &       ---   & $8.15\pm0.15$ & $236\pm5$ & (7) & 147 masers     \\
 $28.63\pm0.26$ &      ---     &      ---      & $7.92\pm0.16$ & $226\pm5$ & (8) & 188 masers     \\
 \hline
 \end{tabular}\end{center}
 {\small\protect (1) Melnik et al. (2001); (2) Zabolotskikh et al. (2002); (3) Bobylev (2017); (4) Rastorguev et al. (2017); (5) Bobylev and Bajkova (2019b); (6) Bobylev and Bajkova (2019a); (7) Reid et al. (2019); (8) Hirota et al. (2020).
 }
  \end{table}
  \begin{table}[t]
  \caption[]{\small
Oort constants $A$ and $B$
 }
  \begin{center}  \label{Oort}    \small
  \begin{tabular}{|c|c|c|c|c|c|c|}\hline
   $A,$     & $B,$     & $\Omega_0=A-B,$ & $V_0,$ & Ref & Sample \\
  km s$^{-1}$ kpc$^{-1}$ & km s$^{-1}$ kpc$^{-1}$ & km s$^{-1}$ kpc$^{-1}$  & km s$^{-1}$  &     &         \\\hline

 $ 16.8\pm0.6$ &           --- &     ---  &    ---   & (1) & Line-of-sight vel. of OB stars \\
 $ 14.4\pm1.2$ & $-12.0\pm2.8$ & $26.4\pm3.0$ & $211\pm24$ & (2) & Various stars \\
 $ 12.9\pm0.7$ & $-16.9\pm1.1$ & $29.8\pm1.3$ & $238\pm11$ & (3) & OB stars     \\
 $ 14.8\pm0.8$ & $-12.4\pm0.8$ & $27.2\pm1.1$ & $218\pm9 $ & (4) & Cepheids, Hipparcos \\
 $ 13.0\pm0.7$ & $-12.1\pm0.7$ & $25.1\pm1.0$ & $201\pm9 $ & (5) & Distant OB stars, Hipparcos \\
 $ 15.9\pm2~~$ & $-16.9\pm2~~$ & $32.8\pm2.8$ & $262\pm23$ & (6) & Red giants, ACT/Tycho-2 \\
 $ 15.3\pm0.4$ & $-11.9\pm0.4$ & $27.2\pm0.6$ & $218\pm6 $ & (7) & Main sequence, Gaia DR1 \\
 $ 15.1\pm0.1$ & $-13.4\pm0.1$ & $28.5\pm0.1$ & $228\pm4 $ & (8) & Main sequence, Gaia DR2 \\
 \hline
 \end{tabular}\end{center}
 {\small\protect (1) Balona and Feast (1973); (2) Kerr and Lynden-Bell (1986); (3) Comeron et al. (1994); (4) Feast and Whitelock (1997); (5) Torra et al. (2000); (6) Olling and Dehnen (2003); (7) Bovy et al. (2017); (8) Li et al. (2019).
 }
 \end{table}

 \subsection*{DISCUSSION}
Tables 2 and 3 give the Galactic rotation parameters derived from various data by analyzing the
angular velocity of rotation, i.e., an approach analogous to ours was used. Some authors prefer to
determine the local Galactic rotation parameters, the Oort constants $A$ and $B.$ Their values are given in
Table 4; from them we calculated the angular velocity and linear rotation velocity of the solar neighborhood
$\Omega_0=A-B$ and $V_0=R_0\Omega_0$ ($R_0=8.0\pm0.15$ kpc).

Tables 2 and 3 present the results obtained by analyzing Cepheids, young open star clusters (OSCs),
OB stars, and masers. Masers are of particular interest. Their trigonometric parallaxes and proper
motions were measured by VLBI with a high (currently
the best) accuracy. Hirota et al. (2020), Reid et al. (2019), and Rastorguev et al. (2017) used data
on masers, which are associated either with very massive O- or B-type protostars or with young low-mass
T Tauri stars. The sources with already measured parallaxes are distributed up to the Galactic center.
Hirota et al. (2020) and Reid et al. (2019) analyze the linear rotation velocity of the Galaxy and, therefore,
they do not provide the derivatives of the angular velocity $\Omega^{\prime}_0$ and $\Omega^{\prime\prime}_0$.

Bobylev and Bajkova (2014) showed that the group velocity components $U_\odot$ and $V_\odot,$ which are
determined from young objects (masers, OB stars, young Cepheids), are noticeably affected by the
perturbations produced by the Galactic spiral density wave. The most reliable value of the Sun’s
peculiar velocity relative to the local standard of rest $(U_\odot,V_\odot,W_\odot)=(11.1,12.2,7.3)$ km s$^{-1}$ is currently believed to have been found by Sch\"onrich et al. (2010).

The group velocity components $U_\odot$ and $W_\odot$ found in the solutions (6) and (7) and given in Table 1 are in good agreement with the values of these parameters
found from other stars (Table 2). In contrast, the
values of the velocity found by us, $V_\odot\sim15$ km s$^{-1}$
(the solutions (6) and (7), Table 1), are typical for very young objects.

The angular velocity of Galactic rotation $\Omega_0$ and its two derivatives found in this paper also suggest
that the stars being analyzed belong to the youngest fraction of the thin disk. The error per unit weight $\sigma_0,$ which is about 16 km s$^{-1}$ in our case, shows a difference.
For example, it is 8--10 km s$^{-1}$ for OB stars and about 14 km s$^{-1}$ for Cepheids. These estimates are
local, i.e., refer to a small solar neighborhood. In our
case, the radius of the solar neighborhood $r=3$ kpc
is fairly large and, therefore, the random errors of
the stellar parallaxes and proper motions contribute
noticeably to the $\sigma_0$ estimate. Note that the error
per unit weight decreases to 12--13 km s$^{-1}$ when
using constraints on the distances to the stars of our
sample, for example, to distances less than 0.5 kpc.
This removes the contradiction related to the large error per unit weight found by us from distant stars.

The linear rotation velocity of the solar neighborhood around the Galactic center $V_0=227\pm4$ km s$^{-1}$
found by us was determined with a high accuracy and is in agreement with the values typical for young
Galactic objects, as can be seen from Tables 3 and 4.

Note the paper by Zari et al. (2018), where more than 40 000 T Tauri stars were selected from the
Gaia DR2 catalogue based on their kinematic characteristics. These stars are no farther than 500 pc
from the Sun. Bobylev (2020) performed a kinematic analysis of the T Tauri stars from the list by Zari
et al. (2018). The error per unit weight $\sigma_0$ for various samples lies within the range 10--12 km s$^{-1}$, while the Oort constants $A$ and $B$ are close to those characteristic
of the Gould Belt, i.e., differ significantly from the Galactic rotation characteristics. In this respect,
the results of our kinematic analysis of a large sample
of T Tauri stars (this paper) at large heliocentric distances are of great interest. An important point is that the Galactic rotation parameters derived from distant low-mass young stars do not differ significantly from those derived from distant high-mass stars (OB stars, masers).

 \subsection*{CONCLUSIONS}
In this paper we used a large sample of young stars located in active star-forming regions to determine
the Galactic rotation parameters. These stars were selected by Marton et al. (2019) based on data from
the Gaia DR2 catalogue by invoking highly accurate photometric infrared measurements from the WISE
and Planck catalogues. We used the stars with a probability of being a YSO more than 80\%.

The distance scale factor was shown to be close to unity and, therefore, the distances calculated via the
Gaia DR2 parallaxes do not require any correction factor. This conclusion is in good agreement with
the analysis of stellar parallaxes from the Gaia DR2 catalogue made by various authors based on various
stars from this catalogue. In addition, we established that the Galactic rotation parameters found are
virtually independent of the linear correction $\Delta\pi=-0.050$ mas to the Gaia DR2 parallaxes.

We considered 25 508 stars within 3 kpc of the Sun with relative trigonometric parallax errors less
than 10\%. T Tauri stars constitute the bulk here. A kinematic analysis using these stars has been performed
for the first time. Based on this sample, we determined the Galactic rotation parameters. In
particular, we obtained a new estimate of the linear rotation velocity of the solar neighborhood around the
Galactic center, $V_0=227\pm4$ km s$^{-1}$. Such a value is typical for the youngest objects of the Galactic thin disk.

 \subsubsection*{ACKNOWLEDGMENTS}
We are grateful to the referee for the useful remarks that contributed to an improvement of the paper.

 \subsubsection*{FUNDING}
This work was supported in part by Program KP19--270 of the Presidium of the Russian Academy
of Sciences ``Questions of the Origin and Evolution of the Universe with the Application of Methods of Ground-Based Observations and Space Research''.

  \bigskip{\bf REFERENCES}{\small

 1. R. Abuter, A. Amorim, N. Baub\"ock, J. P. Berger, H. Bonnet, W. Brandner, Y. Cl\'enet,
V. Coud\'edu Foresto, et al. (Gravity Collab.), Astron. Astrophys. 625, L10 (2019).

 2. R. Adam, P. A. R. Ade, N. Aghanim, M. I. R. Alves, M. Arnaud, M. Ashdown, J. Aumont, C. Baccigalupi,
et al. (Planck Collab.), Astron. Astrophys. 594, 10 (2016).

 3. F. Arenou, X. Luri, C. Babusiaux, C. Fabricius, A. Helmi, T. Muraveva, A. C. Robin, F. Spoto, et al.
(Gaia Collab.), Astron. Astrophys. 616, 17 (2018).

 4. L. A. Balona and M. W. Feast, Mon. Not. R. Astron. Soc. 167, 621 (1973).

 5. A. Blaauw, Ann. Rev. Astron. Astrophys. 2, 213 (1964).

 6. V. V. Bobylev and A. T. Bajkova, Mon. Not. R. Astron. Soc. 441, 142 (2014).

 7. V. V. Bobylev and A. T. Bajkova, Astron. Lett. 41, 473 (2015).

 8. V. V. Bobylev and A. T. Bajkova, Astron. Lett. 44, 184 (2018).

 9. V. V. Bobylev and A. T. Bajkova, Astron. Lett. 45, 20 (2019).

 10. V. V. Bobylev and A. T. Bajkova, Astron. Lett. 45, 331 (2019).

 11. V. V. Bobylev, Astron. Lett. 46, 131 (2020).

 12. J. Bovy, Mon. Not. R. Astron. Soc. 468, L63 (2017).

 13. A. G. A. Brown, A. Vallenari, T. Prusti, J. de Bruijne, F. Mignard, R. Drimmel, C. Babusiaux,
C. A. L. Bailer-Jones, et al. (Gaia Collab.), Astron Astrophys. 595, 2 (2016).

 14. A. G. A. Brown, A. Vallenari, T. Prusti, J. de Bruijne, C. Babusiaux, C. A. L. Bailer-Jones, M. Biermann,
D. W. Evans, et al. (Gaia Collab.), Astron. Astrophys. 616, 1 (2018).

 15. J. Byl and M. W. Ovenden, Astrophys. J. 225, 496 (1978).

 16. T. Camarillo, M. Varun, M. Tyler, and R. Bharat, Publ. Astron. Soc. Pacif. 130, 4101 (2018).

 17. F. Comeron, J. Torra, and A. E. Gomez, Astron. Astrophys. 286, 789 (1994).

 18. R. M. Cutri, E. L. Wright, T. Conrow, J. Bauer, D. Benford, H. Brandenburg, J. Dailey, et al.,
VizieR On-line Data Catalog: II/311 (2013). 
 http://wise2.ipac.caltech.edu/docs/release/allsky/expsup/index.html.

 19. A. K. Dambis, A. M. Mel'nik, and A. S. Rastorguev, Astron. Lett. 27, 58 (2001).

 20. T. Do, A. Hees, A. Ghez, G. D. Martinez, D. S. Chu, S. Jia, S. Sakai, J. R. Lu, et al., Science (Washington, DC, U. S.) 365, 664 (2019).

 21. M. Feast and P. Whitelock, Mon. Not. R. Astron. Soc. 291, 683 (1997).

 22. D. Fern\'andez, F. Figueras, and J. Torra, Astron. Astrophys. 372, 833 (2001).

 23. J. A. Frogel, and R. Stothers, Astron. J. 82, 890 (1977).

 24. Y. M. Georgelin and Y. P. Georgelin, Astron. Astrophys. 49, 57 (1976).

 25. R. de Grijs and G. Bono, Astrophys. J. Suppl. Ser 232, 22 (2017).

 26. T. Hirota, T. Nagayama, M. Honma, Y. Adachi, R. A. Burns, J. O. Chibueze, Y. K. Choi,
K. Hachisuka, et al., arXiv: 2002.03089 (2020).

 27. F. J. Kerr and D. Lynden-Bell,Mon. Not. R. Astron. Soc. 221, 1023 (1986).

 28. C. Li, G. Zhao, and C. Yang, Astrophys. J. 872, 205 (2019).

 29. C. C. Lin, C. Yuan, and F. H. Shu, Astrophys. J. 155, 721 (1969).

 30. L. Lindegren, J. Hernandez, A. Bombrun, S. Klioner, U. Bastian, M. Ramos-Lerate, A. de Torres, H. Steidelmuller, et al. (Gaia Collab.), Astron. Astrophys. 616, 2 (2018).

 31. G. Marton, P. \'Abrah\'am, E. Szegedi-Elek, J. Varga, M. Kun, \'A. K\'osp\'al, E. Varga-Vereb\'elyi, S. Hodgkin, et al., Mon. Not. R. Astron. Soc. 487, 2522 (2019).

 32. A. M. Mel’nik, A. K. Dambis, and A. S. Rastorguev, Astron. Lett. 27, 521 (2001)

 33. A. M. Mel’nik and A. K. Dambis, Astron. Rep. 62, 99 (2018)

34. J. M. Mohr and P. Mayer, Bull. Astron. Inst.
Czechosl. 8, 142 (1957).

35. R. P. Olling and W. Dehnen, Astrophys. J. 599, 275
(2003).

36. J. H. Oort, Bull. Astron. Inst. Netherlands, 3, 2750
(1927).

 37. J. S. Plaskett and J. A. Pearce, Mon. Not. R. Astron. Soc. 69, 80 (1934).

 38. T. Prusti, J. H. J. de Bruijne, A. G. A. Brown, A. Vallenari, C. Babusiaux, C. A. L. Bailer-Jones, U. Bastian, M. Biermann, et al. (Gaia Collab.), Astron. Astrophys. 595, A1 (2016).

 39. A. S. Rastorguev, M. V. Zabolotskikh, A. K. Dambis, N. D. Utkin, V. V. Bobylev, and A. T. Bajkova, Astrophys. Bull. 72, 122 (2017).

 40. M. J. Reid, K. M. Menten, A. Brunthaler, X. W. Zheng, T. Dame, Y. Xu, J. Li, N. Sakai,
et al., Astrophys. J. 885, 131 (2019).

 41. A. G. Riess, S. Casertano, W. Yuan, L. Macri, B. Bucciarelli, M. G. Lattanzi, J.W. MacKenty, J. B. Bowers,
et al., Astrophys. J. 861, 126 (2018).

 42. C. V. Rubin and J. Burley, Astron. J. 69, 80 (1964).

 43. D. Russeil, Astron. Astrophys. 397, 133 (2003).

 44. R. Sch\"onrich, J. Binney, and W. Dehnen, Mon. Not. R. Astron. Soc. 403, 1829 (2010).

 45. J. Torra, D. Fern\'andez, and F. Figueras, Astron. Astrophys. 359, 82 (2000).

 46. J. P. Vall\'ee, Astrophys. Space Sci. 362, 79 (2017).

 47. E. L. Wright, P. R. M. Eisenhardt, A. K. Mainzer, et al., Astrophys. J. 140, 1868 (2010).

 48. Y. Xu, S. B. Bian, M. J. Reid, J. J. Li, B. Zhang, Q. Z. Yan, T.M.Dame, K. M. Menten, et al., Astron.
Astrophys. 616, L15 (2018).

 49. L. N. Yalyalieva, A. A. Chemel’, E. V. Glushkova, A. K. Dambis, and A. D. Klinichev, Astrophys. Bull.
73, 335 (2018).

 50. M. V. Zabolotskikh, A. S. Rastorguev, and A. K. Dambis, Astron. Lett. 28, 454 (2002).

 51. E.Zari, H. Hashemi, A. G. A. Brown, K. Jardine, and P. T. de Zeeuw, Astron. Astrophys. 620, 172 (2018).

 52. P. T. de Zeeuw, R. Hoogerwerf, J. H. J. de Bruijne, A. G. A. Brown, and A. Blaauw, Astron. J. 117, 354
(1999).

 53. J. C. Zinn, M. H. Pinsonneault, D. Huber, and D. Stello, arXiv: 1805.02650 (2018).

  }
  \end{document}